\documentclass[superscriptaddress,eqsecnum,amsfonts,showpacs]{revtex4}
\usepackage{epsfig}
\newcommand{\be}{\begin{equation}}
\newcommand{\ee}{\end{equation}}
\newcommand{\bea}{\begin{eqnarray}}
\newcommand{\eea}{\end{eqnarray}}

\newcommand{\nn}{\nonumber}
\newcommand{\ep}{i\epsilon}
\newcommand{\om}{\omega}

\begin{document}

\title{ The quark spectral functions and the Hadron Vacuum Polarization from application of DSEs in Minkowski space}
\author{V. \v{S}auli}
\affiliation{Department of Theoretical Physics, NPI Rez near Prague, Czech Academy of Sciences. Czech Republic, orcid:0000-0003-3306-840X}
\email{sauli@ujf.cas.cz}

\begin{abstract}
 \begin{center}{\bf Abstract: }\end{center}
The   hadronic vacuum polarization function $\Pi_h$  for two light flavors 
 is computed on the entire domain of spacelike and  timelike momenta using a framework of Dyson-Schwinger equations.
The analytical continuation of the function $\Pi_h$ is based on the utilization of the Gauge Technique with the entry
of QCD Green's functions determined from generalized quark spectral functions. For the first time, 
the light quark spectral functions are extracted from the solution of the  gap equation 
for the quark propagator.  The scale is set up by the phenomena of dynamical chiral symmetry breaking,
 which is a striking feature of low energy QCD. 
\end{abstract}
\pacs{11.10.St, 11.15.Tk}

\maketitle

%%%%%%%%%%%%%%%%%%%%%%%%%%%%%%%%%%%%%%%%%%%%%%%%%%%%%%%%%%%%%%%%%%%%%%%%%%%%%%%%%%%%%

\section{Introduction}

 The hadron vacuum polarization function $\Pi_h(x)$  is conventionally defined through
  the vacuum expectation of current-current correlator such that  $\Pi_h^{ab}(x)=\sum_q <0|j_q^a(0)j_q^b(x)|0>$ where the sum runs over 
 the all quark flavors. It is also an alternative name for the part of the  photon self-energy $\Pi(x)$ due to the quark loops. 
 Using the continuous functional  formalism it can be precisely  defined 
 as double differentiation of  generating functional $\Gamma[\phi_{SM}]$ with respect to the photon fields $A$:
 \be \label{def}
 \Pi(x-y)^{\mu\nu}=\frac{\delta^2\Gamma[\phi_{SM}]}{\delta A^{\mu}(x) \delta A^{\nu}(y)}|_{\phi_{SM}}-
 ... \, ,
 \ee
 where  $\phi_{SM}$ stand for  whole known ensemble of Standard Model fields and where
 the dots stands for the inverse of the free photon propagator.
 Using a standard routine \cite{ITZU1980, ROWI1994} one can derive for the hadronic part of the Fourier transform of (\ref{def}) 
 a well known expression
\be \label{DSEVHP}
\Pi_h^{\mu\nu}(s)=-i e^2 N_c \sum_q e_q^2  Tr \int \frac{d^4k}{(2\pi)^4} \Gamma_q^{\mu}(k-q,k) S_q(k)\gamma^{\nu} S_q(k-q)   \, ,
\ee
 where  the photon momentum satisfies $s=q^2$, $e_q$ is the quark charge in units of electron charge $e$ and the trace is taken in Dirac space and $ \Gamma_q^{\mu}$ is the dressed quark-photon proper vertex,
 $S_q$ is the dressed quark propagator, both functions  satisfy their own Dyson-Schwinger equations,  solutions of them in Minkowski space will be the subject of presented paper.

  Together with the leptonic  polarization function and  loops containing gauge bosons $W,Z$, the function $\Pi_h$  completes the (inverse) photon propagator (\ref{def}). 
   In the spacelike domain of momenta the  polarization function is responsible for  a smooth and slow increase of the  running QED charge. However, 
   for positive $s$ the complexity of hadronic polarization $\Pi_h$   causes measurable interference effect in the fine structure constant  $\alpha_{QED}$.

It is an experimental fact, that  heavier quark $q$ is  a larger quantum fluctuations in the function   $\Pi_h$ one gets.
Thus  at the so called B-factories   like BABAR \cite{psicko}, BESS and BELLE one can easily see  
an   enhancement in muon pair production at vicinity  of bottomonium energies $s\simeq  M_{\Upsilon}^2$ of colliding pair  $e^+e^-$
,the effect for strangeonium  $\phi$ meson energy \cite{phicko} gets substantially smaller,
 while  the  precise KLOE2 experiment  observed such effect bellow 1 GeV energy  \cite{KLOE2} only very recently.
Photon polarization function  offers a great amount of physical information 
and in the timelike domain of momenta, it  is measured with continuously improved accuracy  for many reasons. 
Needless to say, most of  nonperturbative methods available in a market deal with the metric of Euclidean space,
 thus being almost blind when trying to look on the timelike domain of Minkowski space.

 Up to  an asymptotically large spacelike momentum the function$\Pi_h$ is not calculable from perturbation theory.
   Historically, the first nonperturbative extraction of  the function $\Pi_h(s)$   come  from the $e^+e^-\to hadrons$ 
   experiments  due to work of Cabbibo \cite{Cabibbo}.  The  method is based on unitarity and analyticity arguments and  it  does not rely 
   on the underlying QCD/QED dynamics at all. Using nonperturbative methods like  lattice  
    QCD   and the functional approach of Dyson-Schwinger equations
 \cite{GFV2011} the function $\Pi_h$ has been obtained at the Euclidean (spacelike) domain of momenta.
    In order to understand how  QCD resonances emerge in the polarization  function $\Pi_h$ and what is the amount of non-resonance background there,
     one  could employ  nonperturbative methods which can  naturally provide an analytical continuation to the Minkowski space.    
To fill this  gap in our knowledge, we extend the use of Nakanishi Integral Representation  
into the  formalism of Dyson-Schwinger equations and  provide the first, albeit very approximate solution
 for the function $\Pi_h$ in the entire domain  Minkowski space momenta.  It is the first application of this kind of calculation  to QCD and we ignore 
 the important part of dressing of quark-photon  vertex function- the vector mesons are not taken into account. The   later could be taken into account only 
 after when future analytical continuation of higher QCD vertices is performed and known, thus calculation presented in this paper represents only the first  but unavoidable step towards a more  complete calculation.

Remind the reader that  the methods based on utilization of  Nakanishi Integral Representation (spectral representation in case of two point functions) were successfully  applied   in quantum models  without confinement for many years  \cite{KUSIWI1997,saI,saII,saIII,KACA2006,SA2008,KACA2010,SAAD2013,SAPAFR2017}.
 Encouraging results for  the electromagnetic form factors were obtained \cite{CAKA2011} within the formalism as well.
 Here we offer generalization to strong coupling  QCD  describing   how to obtain quark spectral representation, which lead to hadron polarization function $\Pi_h$ with expected \''correct\'' analytical properties.

In the next section a minimal system of $QCD\&QED$ Dyson-Schwinger equations 
is presented. In the Section III, the Gauge Technique is reviewed for the quark propagator satisfying 
Nakanishi Integral Representation and resulting  formula for function $\Pi_h$ is presented there as well.
A novel powerful technique necessary  to get the numerical solution and results   are presented in the last Section IV.

\section{Expressing the hadron vacuum polarization in Minkowski space}

The Dyson-Schwinger equations are an infinite system of quantum equations of motions for Green's functions and
when solved exactly the would provide the full information about theory.
Continuous formalism of Dyson-Schwinger equations has found its most important applications 
in evaluations of hadronic properties within the use of QCD degrees of freedom: quark and gluon fields.   
 Bethe-Salpeter equation is part of the system and   traditional tool for calculation of meson masses \cite{HGKL2017,SAJPSI,SAdva,HPGK2015,DKHK2014},  electromagnetic form-factors \cite{MT99,BPT2003,kacka2005} and meson transition form factors \cite{RCBCGR2016,DRBBCCR2019}.

Here we begin with the QCD part of  the model and restrict to two  flavors QCD $q=u,d$ in the isospin (equal mass) limit.
 The solution for the quark propagator can be represented by  two scalar functions: 
\be \label{cuba}
 S(q)=\not q S_v(q)+ 1 S_s(q)=[A(q)\not q-B(q)]^{-1} \, \, ,
\ee
where the inverse of $A$ is  the renormalization wave function, 
while the (renormalization invariant) quark dynamical mass function is conventionally defined as $M=B/A$.

 In this paper, we begin with the simplest symmetry preserving  
truncation of the equations system - the  Ladder-Rainbow Approximation, for which the kernel  at arbitrary linear gauge $\xi$  is chosen:
\bea \label{potent}
V(q)&=&\gamma_{\mu}\times\gamma_{\nu} \left(-g^{\mu\nu} V_{V}(q)-\frac{4g_L^2}{3}\xi\frac{L^{\mu\nu}(q)}{q^2}\right)   \,\, ,
\\
V_{V}(q)&=&\frac{c_V(m_g^2-\Lambda_g^2)}{(q^2-m_g^2+\ep)(q^2-\Lambda_g^2+\ep)}\, \, ,
\\
L^{\mu\nu}(q)&=&q^{\mu}q^{\nu}/{q^2} \, \, .
\eea
The quark Dyson-Schwinger equation can be written as
\bea   \label{gap}
S^{-1}(q)&=&\not q- m_q-\Sigma(q) \, ,
\nn \\
\Sigma(q)&=&i\int\frac{d^4 k}{(2\pi)^4} \gamma_{\mu} S(k) \gamma^{\mu} V_{V}(k-q) 
\nn \\
&-&i\frac{4\xi g^2}{3}\int\frac{d^4 k}{(2\pi)^4} \gamma_{\mu} S(k) \gamma_{\nu} \frac{L^{\mu\nu}(k-q)}{(k-q)^2}   \, .
\eea
where, as explained in further text, the numerical  values of five parameters appearing  in Eqs. (\ref{potent}),(\ref{gap}) 
are determined by the pion properties:e.g. by the pion mass and the pion decay constant 
with further  requirement that the vacuum hadron polarization function $\Pi_h(s)$ has a cut  at 
the timelike positive axis of $s$, and neither  poles or  branch points and associated cuts are not allowed everywhere else.
The analytical form of  kernel (\ref{potent}) is one of key ingredients for compliance with desired 
analyticity of the function $\Pi_h$.

Another good motivation for the use of the kernel (\ref{potent}) 
is that its  generalization is very straightforward.
Actually, within  the method of Nakanishi Integral Representation
the whole formal derivations presented in this paper remains valid for a large  class of possibly 
considered interactions. To begin with the simplest, we  avoid further integrations and  stay with two poles in the kernel 
characterized by two constant masses $\mu_g$ and  $\Lambda_g$ respectively. 
Nonetheless, kindred Bethe-Salpeter equation (BSE) models  \cite{SAJPSI,SAdva, SApion} turned out to be 
 successful in description of ground  and excited states of pions and charmonia. 

Let us clarify, that the method based on utilization  of Nakanishi Integral Representation we employ here
 can hardly  compete with  impressive amount of achievements already made in the Euclidean space,
 either obtained in the   Rainbow-Ladder Approximation \cite{HGKL2017,SAJPSI,MVRB2017,GKL2018,SEK2017} 
or calculated  with even more  sophisticated truncation \cite{SAWI2018}.  
Instead of, the main goal here, is to provide the first  reliable form of  generalized quark spectral functions.
Within use of them, the function $\Pi_h$  will be obtained 
in the entire domain of the Minkowski space momentum for the first time.

Depending on values of parameters appearing in  the equation (\ref{gap}), we can get  many curious  solutions. 
In order to extract solution, which is  consistent with QCD dynamic one needs to reproduce correct hadron properties, 
e.g.  properties of lightest meson - the pion-.
The meson bound states in the vacuum are described by  (BSE) which explicitly depends on the 
dressed momentum dependent quark propagator, determined by the quark equation (\ref{gap}). For the sake of consistency, 
the BSE and the Eq. (\ref{gap})  must use the identical kernel.
For  this purpose we  solve the pion BSE, which reads:
 \bea \label{BSE}
\Gamma(P,p)&=&i  \int\frac{d^4k}{(2\pi)^4}\gamma_{\mu}S_q(k_+)\Gamma(P,k)S_q(k_-)\gamma_{\nu} [-g^{\mu\nu} V_V(p-k)-4/3\xi g_L^2\frac{L^{\mu\nu}(p-k)}{(p-k)^2}] \, ,
\eea
where P is the total momentum of meson satisfying $P^2=m_{\pi}^2$ for the ground state pion and  the arguments in the quark propagator are $k_{\pm}=k\pm{P/2}$.
 The pion  BSE vertex  function  reads  
\bea
\Gamma(P,p)=\gamma_5\left[\Gamma_A(P,p)+\not p \Gamma_B(P,p)+\not P \Gamma_C(P,p)+[\not p,\not P] \Gamma_D(P,p)\right] \, ,
\eea
where $\Gamma_X$ are four contributing  scalar functions. 

The equation (\ref{BSE}) has been solved by method of iterations described in the papers \cite{SAJPSI,SAdva, SApion}.

In order to get solution of gap equation \ref{gap}, we  have found advantageous to work in a class of linear gauges  and solve
the gap equation in any gauge we wish.

The  Ladder-Rainbow approximated kernel that we use  is often  regarded as   the product of the gluon propagator $G_g$  and the quark-gluon vertex function $\Gamma_{\mu}$ proportional to $\gamma^{\mu}$ matrix, where both function transforms non-trivially when sliding with the  gauge fixing parameter $\xi$. 
In other words, two aforementioned Green's  functions are implicitly gauge dependent and changing the single value of $\xi$ in equations above  does not mean changing the gauge fixing condition properly. Nevertheless, we introduce  the longitudinal parameters  in  gap equation (\ref{gap}) and BSE (\ref{BSE})
and we found its presence is advantageous from various reasons, e.g. from  numerical point of view. In other words, it allows to identify the value of the product $g\xi$ such that  the rest of the model provides a cheap and easy evaluation of  correct pion properties (pion mass and decay width) as well as   
 it allows correct  evaluation  of  hadron vacuum polarization function in the entire Minkowski space.

\subsection{Gauge Technique entry}
\label{gaugen}  
  
  To evaluate Eq. (\ref{DSEVHP}) one needs to know the  solution for the Abelian gauge vertex $\Gamma^{\mu}$.
The best would be to solve the Dyson-Schwinger equation for this vertex as well, however to do this in Minkowski space is
recently impossible. To get first reliable result in off resonance region, we appreciate the fact that $U(1)$ 
electromagnetic symmetry is unbroken in the Nature and employ the Gauge Technique. This  allows us 
 to close the system of DSEs  by construction of the quark-photon vertex in a minimal gauge invariant manner.

 For practical purpose we will use un-amputated vertex, which relates  the propagators and the proper gauge  vertex in  usual way:
\be
\Lambda^{\mu}(p,l)=S(p)\Gamma^{\mu}(p,l)S(l)
\ee
and solve the Ward-Takahashi identity, which reads 
\be
(p-l)^{\mu}\Lambda_{\mu}(p,l)=S(p)-S(l)
\ee 

The Gauge Technique  has been  introduced in Ref. \cite{DEWE1977} and it represents gauge covariant tool for solution of 
 Dyson-Schwinger equation in the entire domain of momenta.
 It consists of writing  a  solution  for  the  vector  un-amputated  vertex  in  the
form
\be
\Lambda^{\mu}(p,q)=\int_{\Gamma} d x \rho(x) \frac{1}{\not p-x}\gamma_{\mu}\frac{1}{\not q-x} \, ,
\ee
where one assumes there exists a generalized spectral representation
for the quark propagator
\be \label{spectral}
S(p)=\int_{\Gamma} d x \frac{\rho(x)}{(\not p-x)} \, .
\ee

In this paper we   generalized Gauge Technique,  which 
instead of spectral representation allows to use  the Hilbert transformation  
\bea \label{dudova}
S_q(p^2)&=&\int_{-\infty}^{\infty} d a \frac{\not p \sigma_v(a)+\sigma_s (a) }{p^2-a+\ep} \, \, .
\eea

The procedure  is relatively  straightforward and the Gauge technique solution for the quark-photon vertex reads:
\bea  \label{gt}
\Lambda^{\mu}(p,q)&=&\int_{-\infty}^{\infty} d \om \frac{ \sigma_v (\om)
 [\gamma^{\mu}\om+\not p \gamma _{\mu} \not q]}{(p^2-\om+\ep)(q^2-\om+\ep)}
\nn \\
&+&\int_{-\infty}^{\infty} d \om \frac{ \sigma_s (\om) [\not p \gamma^{\mu}+ \gamma _{\mu} \not q]}{(p^2-\om+\ep)(q^2-\om+\ep)}
+\Lambda^{\mu}_T(p,q) \, ,
\eea
where the  transverse piece satisfies $Q.\Gamma_T(p,q)=0$ and  where $Q$ is   photon momentum.
 Inclusion of the transverse components  $\Lambda^{\mu}_T(p,q)$ requires the solution of Dyson-Schwinger equation for the quark-photon vertex and 
 is subject of recent study \cite{recent}. By converting the momentum space gap equation for the vertex into a new but equivalent 
integro-differential equation for the so called Nakanishi weights, it  turns out to be feasible task, 
which could provide the solution in the entire  Minkowski space. 
Leaving this important calculation for a future work, we take  $\Lambda^{\mu}_T(p,q)=0$ in this paper for purpose of simplicity.

 The Eq. (\ref{dudova}) should be regarded as the generalization of Lehmann representation, 
 with two properties in absence: neither of function $\sigma_v$ or $\sigma_s$ 
is positive definite and the position of branch point is not assumed in advance. 
The introduction of  negative  cuts in  relations (\ref{dudova}) and (\ref{gt}) could be regarded as  
 as an  auxiliary  step, which when missing, the solution of Dyson-Schwinger equation for the quark propagator 
 would be hardly achievable in practice, if  possible at all.
 Anticipate here the solution: the negative cut is gradually vanishing as it is subject of minimization.
Furthermore, no prohibited acausal behavior is observed as a solution. 
There is no evidence for negative branch point associated 
with appearance of pathological singularities like tachyonic poles. 

  In fact,  we do not expect and we actually do not get the quark propagator  pole within the timelike axis as well.
 The observed absence of the real pole in the propagator  is  the analytical realization of  
 confinement mechanism in presented framework. At last but not at least, let us make a  technical note:  
 the so called Wick rotation contour can be used for purpose of   analytical continuation  into  the Euclidean space.
 This   fact will be silently used during the derivation.

Using the Gauge technique (\ref{gt}), the relation (\ref{DSEVHP}) necessary to evaluate reads
\be  \label{must}
\hat{\Pi}_h(q)=\Sigma_f \frac{e^2_f N_c}{3}Tr_D\int\frac{d^4k}{(2\pi)^4}\Lambda^{\mu}(p-q,q)\gamma_{\mu} \, ,
\ee 
which within the use of  on-shell renormalization prescription and after long but rather  straightforward calculation
  gives the desired result:
\bea  \label{hoho}
\hat{\Pi}(q)&=&q^2{\Pi}(q) \, ;
\nn \\
\Pi_h(q)&=&\Sigma_{f=u,d} \frac{e^2_q N_c}{4\pi^2} 
\frac{8 q^2}{3} \int_0^1 dx X(x) S_v^q (a) 
\eea
where the argument of the function $S_v$ (see Eq. \ref{cuba})  is $a=x(1-x)q^2$ and
\be
X(x)=4x^4+3x^3-x^2 \, .
\ee

The expression  (\ref{hoho}) shows an elusive  way how the dressed quark propagator appears
in the  hadron polarization  function $\Pi_h$ . As the Eq. (\ref{hoho}) is deduced from  gauge covariant 
consideration,  this term should be always presented in  any other meaningful approximation.

\section{Results, HVP constrained by the  pion properties and vice versa}

We got the correct physical picture within the model and reproduce
the static properties of the pion i.e. pion mass $m_{\pi}=140MeV$,  pion decay constant $f_{\pi}\simeq 95MeV$ as well as we reproduced desired analytical properties of continuous  form factors, i.e. the  function $\Pi_h$ in our case.
In order to get static property, the quark propagator was substituted into the BSE (\ref{BSE}) in order to identify the pion bound state.
For this purpose we accommodate the method developed for kindred model   described in the paper \cite{SAJPSI,SAdva}.
The method works  in its Euclidean approximation,  it requires the quark propagator evaluated  at complex value of momenta, 
which was  quite  easily achieved within the use  of  the integral representation (\ref{dudova}).  
As a consequence, the solution of BSE stays  completely real in isospin limit considered here.

To get the solution for the function $\Pi_h$ one just needs to substitute the quark propagator $S$ into the expression (\ref{hoho})
and integrate over the variable $x$ , which was done numerically.  To get the propagator $S$ we converted the Eq. (\ref{gap}) into the 
 integral equations for  functions $\sigma_v$ and $\sigma_s$ and solve it by the method of iterations.
As we can get the quark spectral function bounded from bellow, we automatically reproduce  the   vacuum hadron  polarization function with the same property. It's absorptive part is exactly zero for the spacelike momenta $q$ in the Eq. (\ref{hoho}). The first solution of this kind has been achieved by complicated brute-force minimization procedure  as described in previous version of this paper. Now and  much easily, the same  result has been  achieved by the method of  momentum subtraction at the timelike scale. The resulting method  is quite simple, it is applicable beyond approximations  made in this paper and  it represents one of the  major achievements  presented in this paper. In details it is described  in the appendix \ref{heureka}, where it is shown it provides the solution for a large region of gauge fixing parameters $\xi$.

\begin{figure}
\centerline{\includegraphics[width=8.0cm]{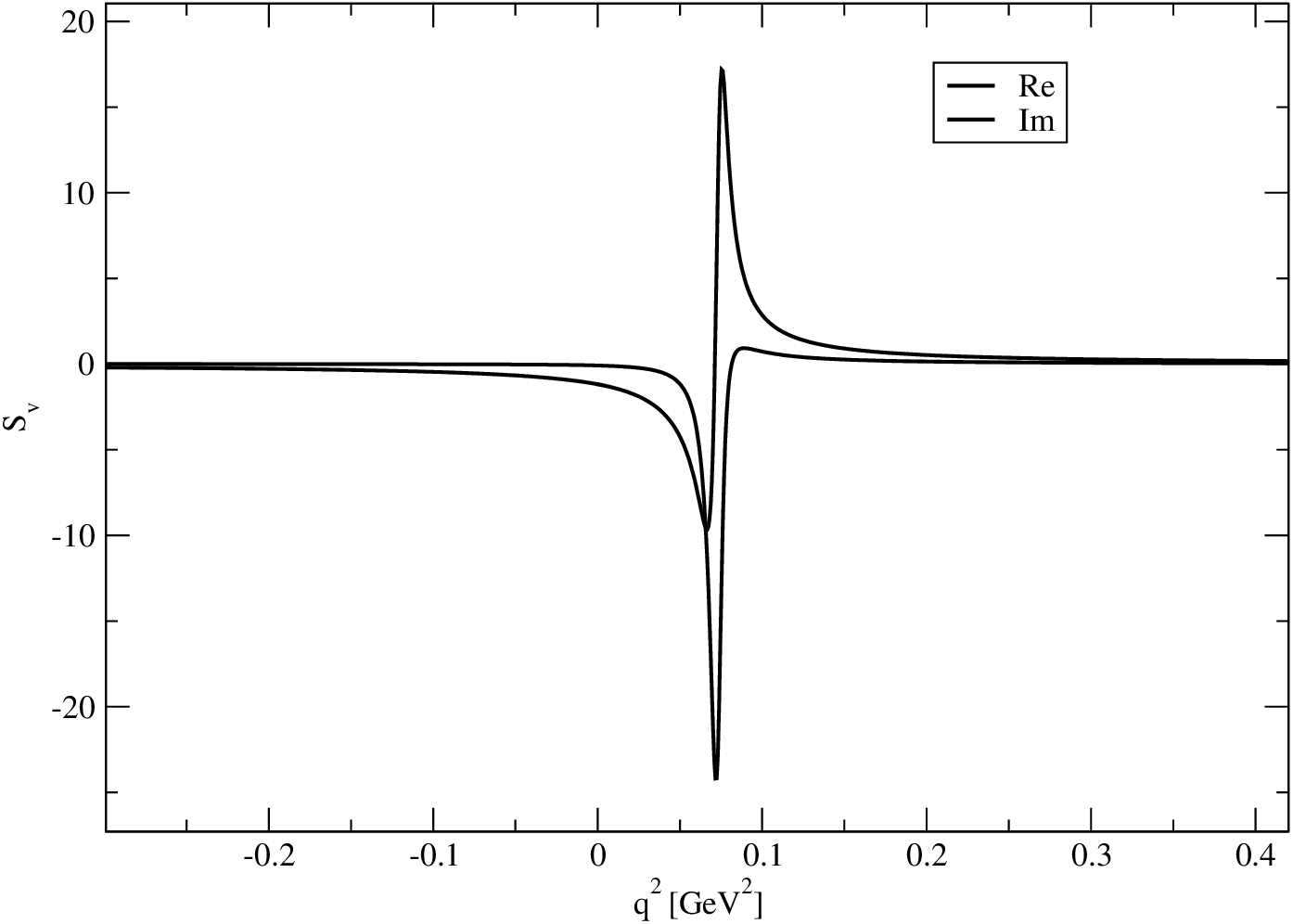}}
\caption{ The quark propagator function $S_v$}
\label{kocka}
{\mbox{-------------------------------------------------------------------------------------}}
\end{figure}

\begin{figure}
\centerline{\includegraphics[width=8.0cm]{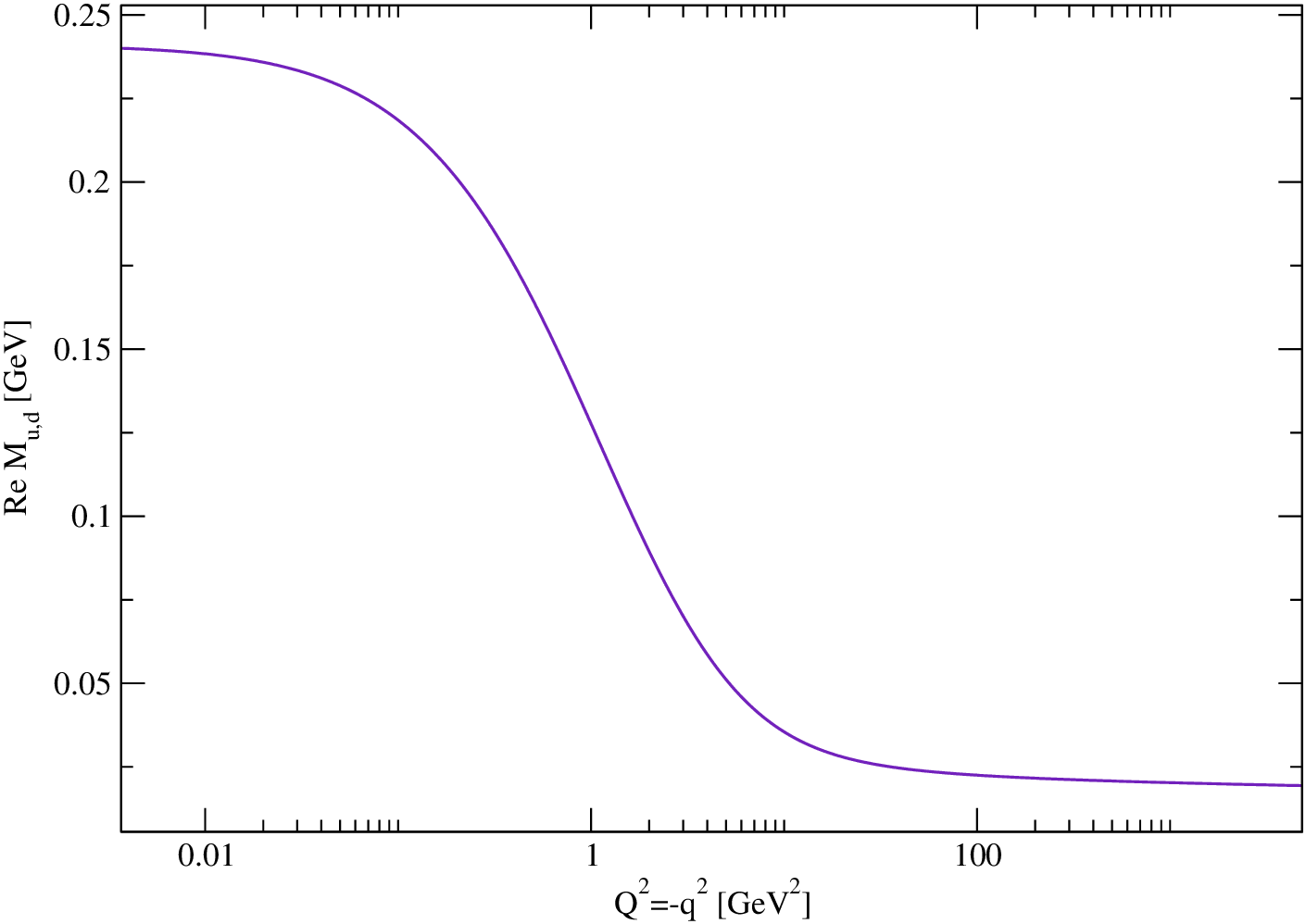}}
\caption{ Typical look known from the Euclidean studies: The quark mass function in the spacelike domain of momenta.}
\label{kocka2}
{\mbox{-------------------------------------------------------------------------------------}}
\end{figure}

\begin{figure}
\centerline{\includegraphics[width=8.0cm]{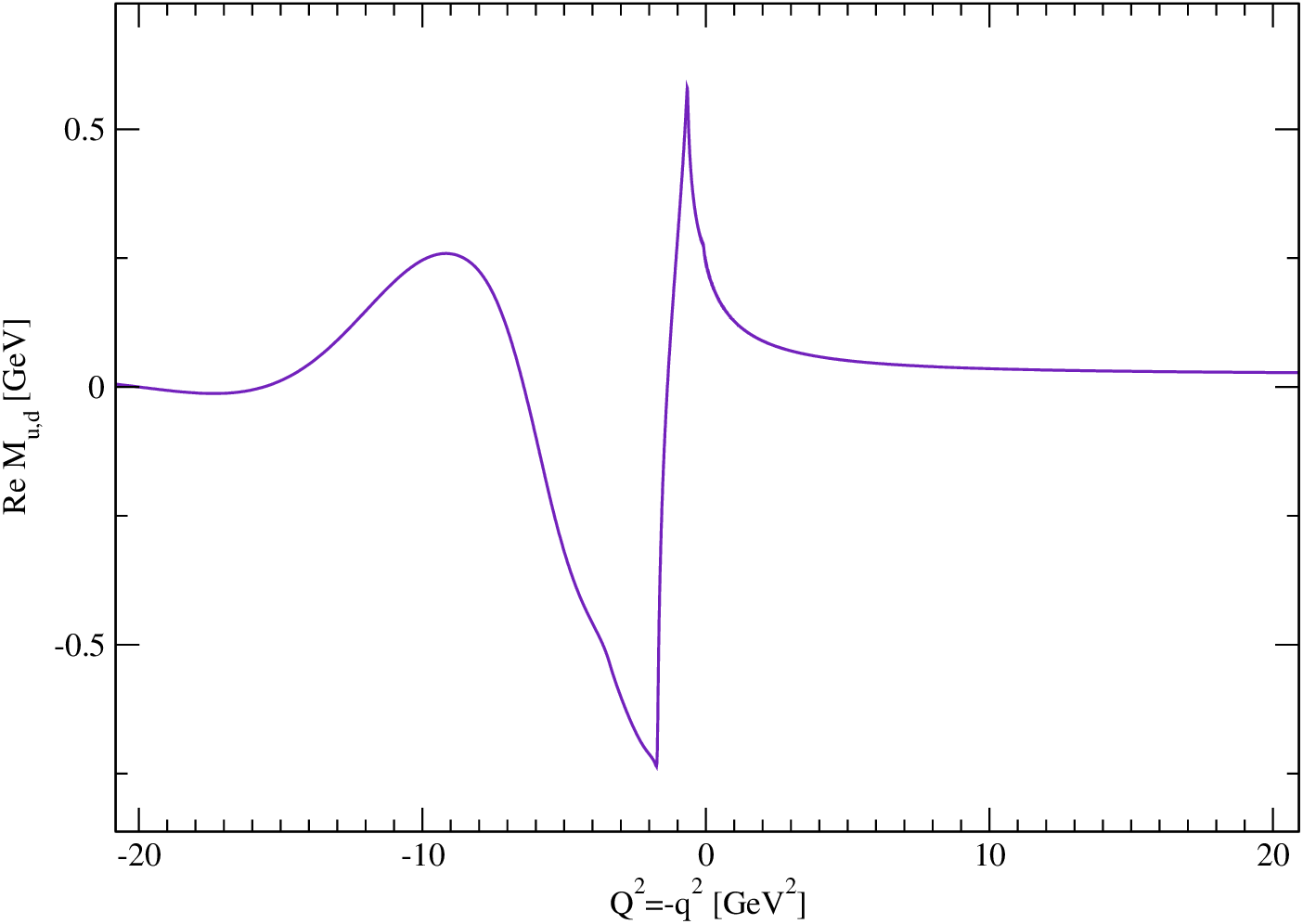}}
\caption{ The same as in previous figure, but larger piece of Minkowski space shown. }
\label{kocka3}
{\mbox{-------------------------------------------------------------------------------------}}
{\mbox{-------------------------------------------------------------------------------------}}
\end{figure}

\begin{figure}
\centerline{\includegraphics[width=8.0cm]{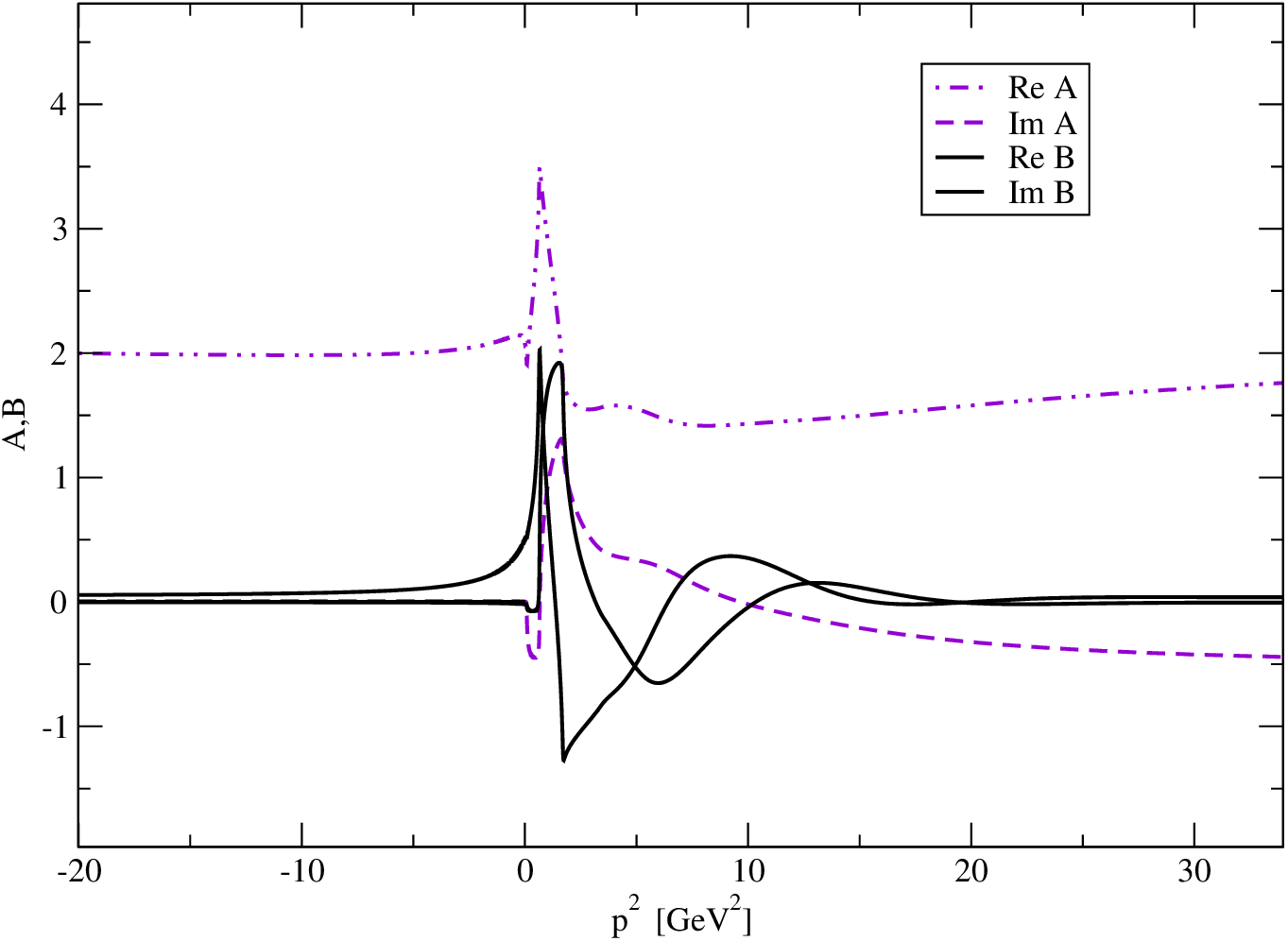}}
\caption{ Conventional look of the quark function $A$ and $B$ in Minkowski space, the  timelike domain of  momenta is on the right side from the origin. }
\label{kocka4}
\end{figure}

 The resulting propagator function, its inverse given by  functions $A,B$  as well as the quark dynamical function $M=B/A$ 
are shown in figures \ref{kocka},\ref{kocka2},\ref{kocka3} and \ref{kocka4} respectively.  Both propagator Nakanishi weights 
change the sign and  also according to confinement phenomena , the quark propagator does not have a real pole.

\begin{figure}
\centerline{\includegraphics[width=9.0cm]{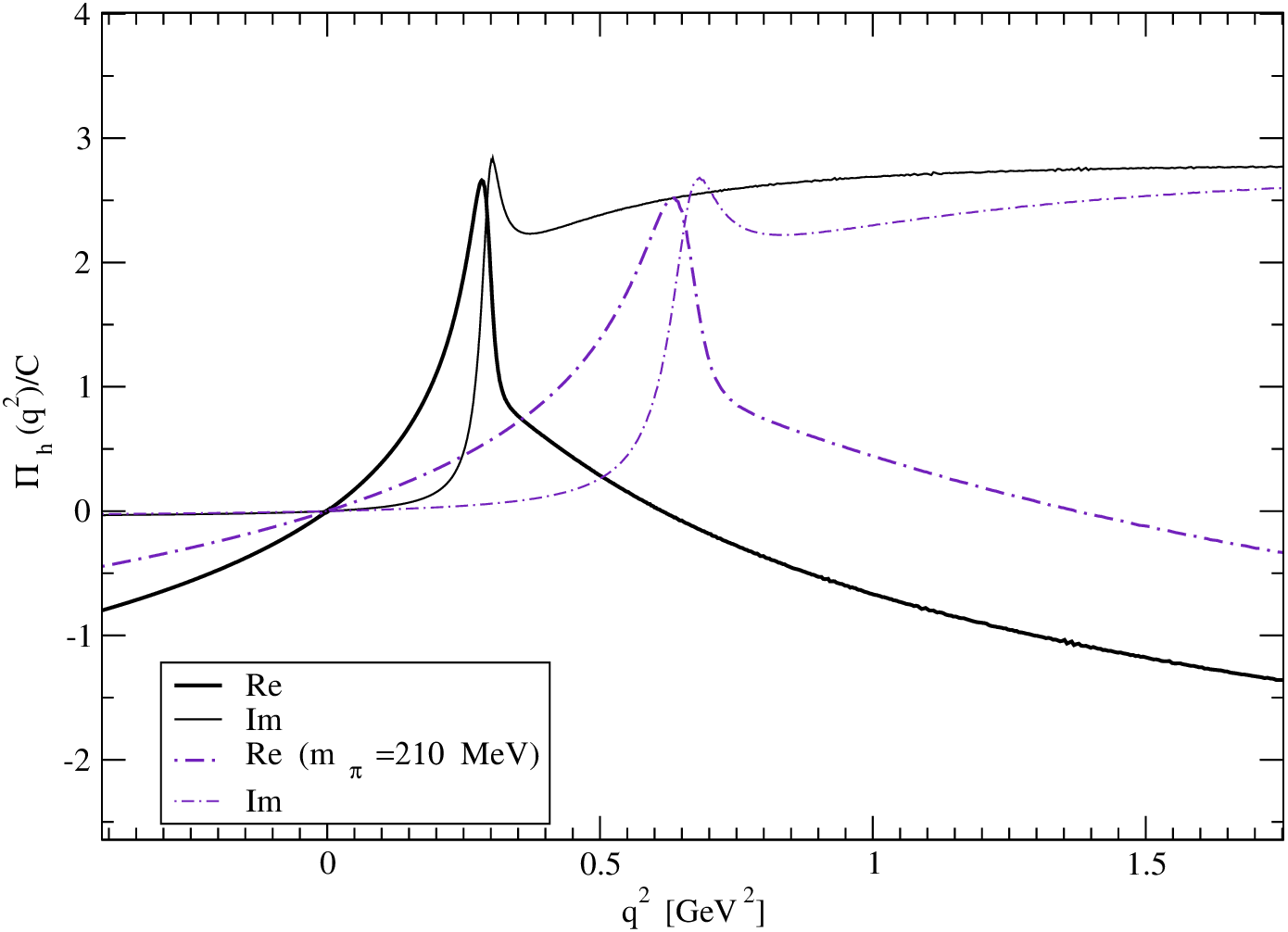}}
\caption{The function $\Pi_h/C$ obtained via Gauge Technique with the constant $C$  defined as  $C=-40\alpha/(9\pi)$. Two distinct curves represent -up to the scale- identical  solution  of Dyson-Schwinger equation.
The  one has  $m_{\pi}=140 MeV$ (solid one) and rescaled one corresponds with $m_{\pi}=210 MeV $.  }
\label{space}
{\mbox{-------------------------------------------------------------------------------------}}
\

\end{figure}

\begin{figure}
\centerline{\includegraphics[width=9.0cm]{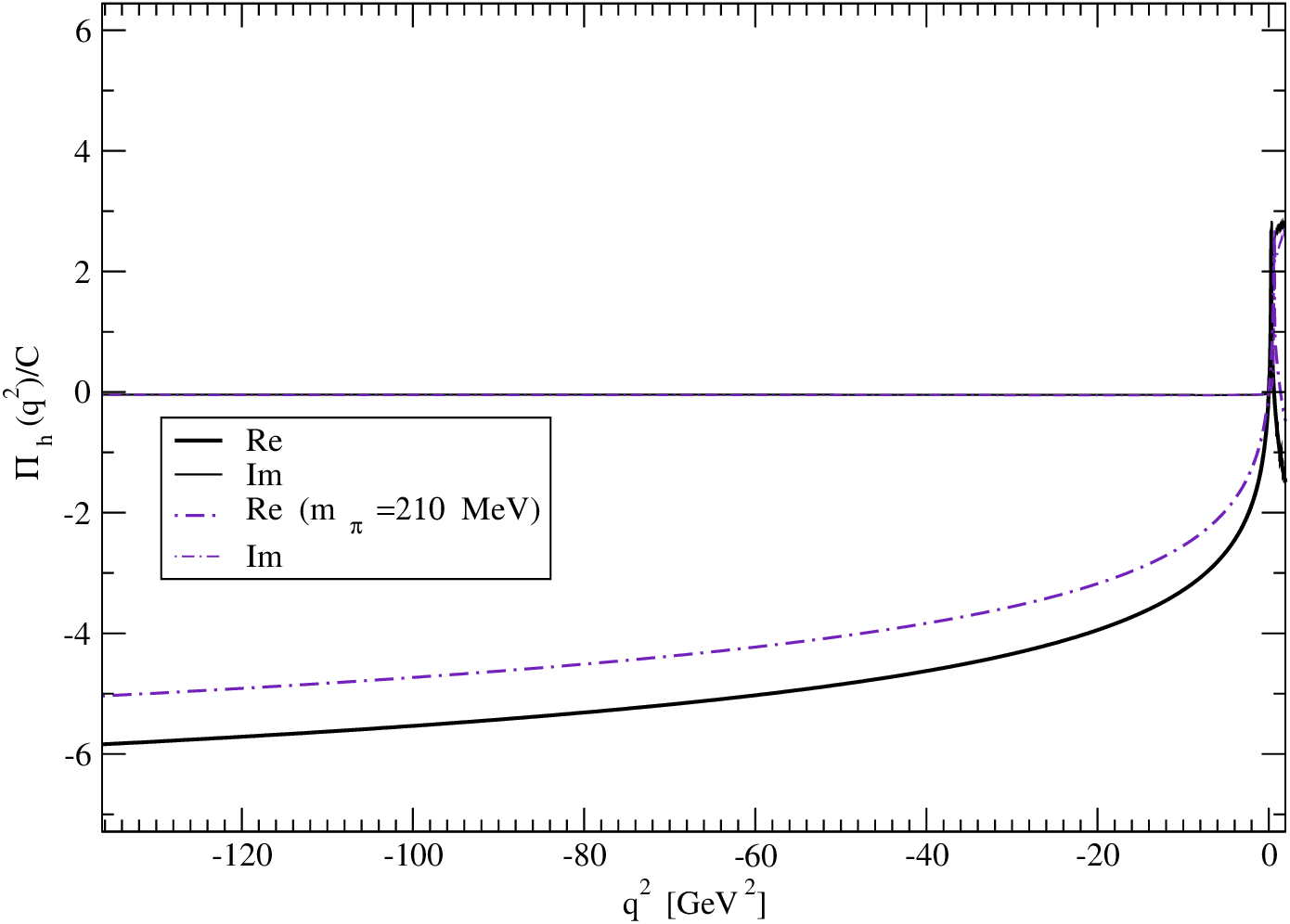}}
\caption{The same as in the previous figure, but zoomed in the spacelike domain of momenta.}
\label{figa}
\end{figure}

If needed one can  add the Eq. (\ref{hoho}) into the coupled set of DSE and BSE and solve the whole system  simultaneously together.
The iterations of all equations were repeated till the model parameters were fitted accordingly and we provide 
 the most important codes for public \cite{gemma}.  Resulting function $\Pi_h$ was calculated within the following parameters  $c_V/(4\pi)^2=1.8$ and $g^2\xi/ (4\pi)^2=0.17$ and $m_g^2/\Lambda^2_g=2/7.5$ ,$m_{\pi}/m_g=1.38/\sqrt{2}$ 
 and is  depicted in  figures \ref{space} and \ref{figa} respectively.  
 
 Due to the absence of transverse  components of the quark-photon vertex, the hadron vacuum polarization  function does not exhibit usual $\rho$ meson peak 
but rather small bump positioned numerically  at $\sqrt{s}= 500 MeV$. This structure  is a certain remnant of the  threshold cusp, which would be presented  when there was a quark propagator  pole in the theory.    Assuming for any reason, that the observed  cusp/bump  in the function $\Pi_h$
 could be hidden  at physical mass of $\rho$ and $\omega$ mesons, one can find another solution of the system by simple rescaling. In this case 
one gets slightly wrong pion mass $m_{\pi}=210$.

\

\

\

\section{Conclusions and Prospects}

We completed a computation of the hadron vacuum polarization function $\Pi_h(q^2)$ in two flavor QCD. All required elements are determined by the solution of QCD's Dyson-Schwinger equations obtained in the rainbow-ladder truncation, the leading order in a systematic and symmetry preserving approximation scheme. Within  novel analysis technique we developed here it was  possible to compute $\Pi_h(q^2)$ on the entire domain of spacelike and  timelike momentum for the first time. 
 
Our calculation  agrees with other methods in the spacelike domain, while at timelike domain  the absence of expected  resonances,
exhibits an expected   fact that the observed bump cannot supply an important pieces 
of the quark-photon vertex we have  not yet included. 
 There is a certain prospect in this direction \cite{recent} and writing the code for its  numerical search is in the progress.

To extract the spectral functions of the quarks a new  method was developed and it provides 
 spectral function , which is in  qualitative accordance with the confinement of  quarks in standard model vacuum.
Including Yang-Mills part more seriously into the game and getting the similar solution 
for  the gluon propagator is  challenging and still opened task for years \cite{CO1982} as well.

\appendix

\section{Rainbow ladder quark self-energy in arbitrary linear gauge}

The Dyson-Schwinger equation for the quark propagator can be converted into the Unitary equations for Nakanishi weights, by comparing  of imaginary and real parts of assumed integral representation (\ref{spectral}) , i.e.   
\be \label{unit2}
\sigma_{v,s}(p^2)=-\frac{\Im S_{v,s}(p^2)}{\pi}  \,\,\,\,\,\,\,\,\,\,\,\,\,  \Re S_{v,s}(p^2)= P.\int ds \frac{\sigma_{v,s}(x)}{p^2-x}  \, ,
\ee
and by the integral representation for the inverse of the propagator, which is  readily derivable  from quark gap equation (\ref{gap}). After the renormalization it reads
\be \label{gap2}
S^{-1}=\not p -m(\mu)-\Sigma(p) \, ,
\ee
 where  the self-energy functions    $\Sigma=\Sigma_{V}+\Sigma_{\xi}$  are evaluated in details in this Appendix.

Let us start with unrenormalized self-energy (hence index 0), which comes from the product of a gauge term and 
the quark propagator expressed through the Hilbert transformation. The first line in (\ref{gap}) reads
\be
\Sigma_{\xi}^0(q)=-i\xi g^2\int \frac{d^4k}{(2\pi)^4} \int d o \, \gamma_{\mu} \frac{\sigma_v(o)\not k+\sigma_s(o)}{(k^2-o+\ep)}\gamma_{\nu}
\frac{(k-q)^{\mu}(k-q)^{\nu}}{{((k-q)^2+\ep)}^2}\, .
\ee 
After a standard treatment and a little algebra it can be written into the following form
\be
\Sigma_{\xi}^0(q)=-\xi g^2\int_0^1 dx \int \frac{d^4 k_E}{(2\pi)^4}\int d o \left[ \frac{4(1-x)\sigma_v(o) k.q \not k}{D^3}
 +\frac{\sigma_v(o)(-2-x) \not q+\sigma_s(o)}{D^2}+ \frac{4(1-x)x^2\sigma_v(o) q^2 \not q}{D^3}\right] \, ,
\ee
where the denominator  $D=-k^2_E-q_E^2(1-x)x-o x$ is strictly negative, noting the Wick rotation is working for  positive as well as for the   negative variable  $o$.
Thus to go to the Euclidean space is what we only need here in order to integrate over the momenta, as the result  we get
\bea \label{haf}
\Sigma_{\xi}^0(q)&=&-\frac{\xi g^2}{(4\pi)^2}\int_0^1 dx \int d o 
((1-x)\not q\sigma_v(o)\left[c(d)+\ln{(\Omega/\mu^2)}\right]
\nn \\
&+&[\sigma_v(o)(-2-x) \not q+\sigma_s(o)][c(d)+\ln{(\Omega/\mu^2)}]
+\frac{(1-x)x \sigma_v(o) q^2 \not q}{q^2(1-x)-o+\ep} \, ,
\eea
where $ \Omega=q^2x(1-x)-o x +\ep$ and $\mu$ is the spacelike renormalization scale 
(  $\mu^2<0$ in our metric convention).

The third term  in the  Eq. (\ref{haf}) is  UV finite and can be rewritten as
\be \label{thd}
-\frac{\xi g^2}{(4\pi)^2}\int_0^1 dx \int d o \frac{(1-x)x \sigma_v(o) q^2 \not q}{q^2(1-x)-o+\ep}
=-\frac{\xi g^2}{(4\pi)^2}\int_0^1 dx \int d o (1-x)x S_v(\om^{,}) q^2 \not q \, ,
\ee
where we have employed the Hilbert transformation (\ref{dudova}) and label $\om^{,}=q^2(1-x)$.

In addition to the usual Minimal Subtraction counter-terms we will sent also the following terms 
\bea \label{picmund}
\delta Z_{\psi}&=&\int_0^1 dx\frac{g^2\xi}{(4\pi)^2}\int d o \sigma_v(o) (1+2x) ln x =\frac{3g^2\xi}{2(4\pi)^2}\int d o \sigma_v(o) 
\nn \\
\delta Z_{m}&=&-\int_0^1 dx\frac{g^2\xi}{(4\pi)^2}\int d o \sigma_s(o)  ln x =-\frac{g^2\xi}{(4\pi)^2}\int d o \sigma_s(o)
\eea
into the renormalization constant $Z_2(Z_{\psi})$ and  $Z_4(Z_{m}) $ .

For the first  two terms in (\ref{haf}) we thus have
\be \label{haf2}
-\frac{\xi g^2}{(4\pi)^2}\int_0^1 dx \int d o \left[ \not q (-1-2x)\sigma_v(o)+\sigma_s(o)] ln \frac{q^2(1-x)-o+\ep}{\mu^2}\right] \, ,
\ee
where we have drop out all renormalization constants.
Using per-partes integration and sending momentum independent  boundary terms into renormalization constants again  one finally gets
for the rest of (\ref{haf2}) : 
\be
q^2 \frac{\xi g^2}{(4\pi)^2}\int_0^1 dx \int d o  \frac{-\not q x(1+x)\sigma_v(o)+x \sigma_s(o)}{q^2(1-x)-o+\ep}
\ee

Summing this  with (\ref{thd} )one finally gets 
\bea \label{third}
\Sigma_{\xi}(q)&=&q^2 \frac{\xi g^2}{(4\pi)^2}\int_0^1 dx \int d o \frac{-2x \sigma_v(o) \not q+x\sigma_s(o)}{q^2(1-x)-o+\ep}
\nn \\
&=&q^2\frac{\xi g^2}{(4\pi)^2}\int_0^1 dx x  [ -2 S_v(\om^{,}) \not q+S_s(\om^{,})]\, .
\eea 
where we have employed the  Hilbert transformation once again.  To end,
we make the substitution $s=\om^{'}$ and  arrive into the desired renormalized result:
\be \label{gauge}
\Sigma_{\xi}(q)=-\frac{\xi g^2}{(4\pi)^2}\int_0^{q^2} d s (1-\frac{s}{q^2})[\not q 2 S_v(s)-S_s(s)]
+C_A \not q +C_B 
\ee
where $C_{A,B}$ are in principle arbitrary finite constants specifying a given renormalization scheme. In this paper we will use modified
 $MS$ for which we take  $C_A=C_B=0$(i.e. the renormalization follow the derivation above).  Fully equivalently we employ momentum renormalization scheme, which defines
 self-energy at given renormalization scheme with renormalization constants  given by renormalized self-energy at given point.
 Numerically , bot methods work equally.  
 
The expression (\ref{gauge} ) is    valid either  for the timelike or the  spacelike momenta $q^2$,
 however for the spacelike arguments $q^2<0$  one should keep in mind 
that the integration runs over the negative values of variable $s$. Thus for an aesthetic reasoning, for spacelike $q$,  we rather write
\be  \label{vus}
\Sigma_{\xi}(q,q^2<0)=\frac{\xi g^2}{(4\pi)^2}\int^0_{q^2} d s \left(1-\frac{s}{q^2}\right)\left[\not q 2 S_v(s)-S_s(s)\right]
+C_A \not q +C_B     \, ,
 \ee
which keeps the   lower integral boundary  smaller then the upper one.

Let us make short digression here and remind  that the dynamical symmetry breaking and massless pion are attributed to chiral limit $m=0$ in QCD. 
This  is a more complicated issue when one is dealing with spectral representation  and as we have sent (irrespective of their UV finitness) scalar 
piece of the self-energy into the renormalized constant, we cannot use our scheme directly for the 
calculation  in the chiral limit. 
To set the mass exactly to zero  one  must also require
\be \label{conda}
\int d o \rho_s(o)=0 \, ,
\ee
which, at least at the  formal level allows us to skip the  mass renormalization at all.
 The sum rule condition (\ref{conda}) could   be explicitly used before the momentum integration in the chiral limit.
otherwise we are  facing the  ambiguity $c(d)\int \sigma_s=\infty .0$ and the result turns to be  ordering dependent (not well defined). 
In this paper we will deal with the physical pion and 
we leave the question of solution in exact chiral limit $m=0$ unanswered for future task. 

 For  the combination of the 
vector interaction $V_v$ with the spectral part of the quark propagator we get
\be
\Sigma_{V}=-i\int\frac{d^4k}{(2\pi)^4} \int d o \frac{\gamma^{\mu}(\not k \sigma_v(o)+\sigma_s(o))\gamma_{\mu}}{k^2-o+\ep}\left[\frac{c_V}{(k-q)^2-m_g^2+\ep}
-\frac{c_V}{(k-q)^2-\Lambda_g^2+\ep} \right] \, ,
\ee
which is  the standard one loop expression  integrated over the continuous mass $o$ giving us the known result: 
\be  \label{nido}
\Sigma_{V}=c_v \int^1_0 dx \int d o \frac{-2 \not q (1-x) \sigma_v(o)+4 \sigma_s(o)}{(4\pi)^2}
log\left(\frac{q^2(1-x)-o-m_g^2\frac{1-x}{x}+\ep}{q^2(1-x)-o-\Lambda_g^2\frac{1-x}{x}+\ep}\right)\, .
\ee

For numerical purpose it is  suited to further proceed by per partes integration
\bea
\Sigma_{V}&=&\frac{c_v}{(4\pi)^2} \int^1_0 dx \int d o \frac{2 \not q (1-x/2) \sigma_v(o)-4 \sigma(o)}
{q^2(1-x)-o-m_g^2\frac{1-x}{x}+\ep}(-q^2+\frac{m_g^2}{x})-(m_g\rightarrow \Lambda_g)
\nn \\
&=&\frac{c_v}{(4\pi)^2} \int^1_0 dx [2 \not q (1-x/2) S_v(\hat{a})-4 S_s(\hat{a})]
[-q^2+\frac{m_g^2}{x}]-(m_g\rightarrow \Lambda_g)\, ,
\eea
 where the argument in  the first term of the second line reads $\hat{a}=q^2(1-x)-m_g^2\frac{1-x}{x}$, which can be seen  by virtue of Hilbert transformation again.

The last step advantageous for numerical solution is the introduction of
the following functional identities 
\be
1=\int_{-\infty}^{\infty} da \delta(a-\hat{a}) \, , \delta(f(x))=\Sigma_i \frac{\delta(x-x_i)}{|\frac{df}{dx}(x_i)|} 
\ee
in the previous equation. Interchanging the order of integration and integrating over the variable $x$ one gets 
\bea  \label{VH}
\Sigma_{V}=\Sigma_{i=\pm}\frac{c_v}{(4\pi)^2}\int\nolimits_{-\infty}^{\infty} da \frac{2 \not q (x_i-x_i/2) S_v(a)-4 x_i S_s(a)}
{sgn(-q^2 x_i^2+m_g^2)}\Theta(x_i)\Theta(1-x_i)\Theta(D)-(m_g\rightarrow \Lambda_g)\, ,
\eea
where the roots are 
\bea
x_{\pm}&=&\frac{-(m_g^2+q^2-a)\pm\sqrt{D}}{-2q^2}
\nn \\
D&=&(m_g^2+q^2-a)^2-4q^2m_g^2
\eea
for the first term. The  expression (\ref{VH}) has been actually used in our numerical code.
 
\section{Determination of quark spectral function}
\label{heureka}

In the Appendix above we have determined the selfenergy functions selfconsistently, i.e. 
assuming the quark propagator satisfies spectral representation 

\bea 
S(k)=\int_0^{\infty}d a\frac{\not p \rho_v(a)+\rho_s(a)}{p^2-a+\ep} \,
\eea   

we have derived  the selfenergy $\Sigma$ , which solely depends on  spectral quark functions $\sigma_v$ and $\sigma_s$.
 
 In addition we assume that the spectral functions are smooth, the only singularity of the  propagator is a cut  at  timelike real axis of momenta, i.e. it starts
 at $p^2=0$. Thus  as a consequence, the propagator is real for the  spacelike arguments $p^2$, $p^2>0$.

Analytically properties  are completely exploited in practice in the numerical search. From this it follows, that for a non-trivial solution for propagator evaluated at timelike  momentum $p^2$, there must be a  unique ratio $\Re S_{s,v}(p^2)/ \Im S_{s,v}(p^2)$ such that $\Im \Sigma_{s,v}(p^2)=-\pi\sigma_{s,v}(p^2)$.

First let us make a projections , writing $S^{-1}(p)=\not p A(p)-B(p)$ where $A,B$ are two scalar function needed to determine the propagator $S$.
Comparing to DSE one gets
\be
B(p^2)= m(\mu)+\Sigma_B(p^2) \, \, ;\, \, A(p^2)= 1-\Sigma_A(p^2)
\ee
where all terms of the selfenergy which are proportional to 4times4times unit matrix are collected in $\Sigma_B(p)$ and all terms 
of selfenergy terms listed  in previous Appendix and proportional to $\not p$ matrix  are collected in the scalar function $\Sigma_A(p)$  
 
In what follows we subtract the two  equations with itself at some arbitrary timelike scale $\zeta$.
\be \label{fuc}
B(p^2)= B(\zeta) +\Sigma_B(p^2)-\Sigma_B(\zeta)
\nn \\
 A(p^2)=A(\zeta)-\Sigma_A(p^2)+\Sigma_A(\zeta)\, .
\ee

To get the spectral functions one just needs to inverse the propagator expressed in terms self-energies.
From the imaginary parts one gets  equation for spectral functions:
\be \label{zdar}
\rho_s(s)=\frac{-1}{\pi}\frac{\Im B(s) R_D(s)+\Re B(s)) I_D(s)}{R_D^2(s)+I_D^2(s)}    
\ee
where $s=p^2>0$ in our metric, and where the functions  $R_D$ and $I_D$  stand for  the square of the real and the imaginary part of the function $sA^2(s)-B^2(s)$, i.e. 
\bea
R_D(s)&=&s[\Re A(s)]^2-s [\Im A(s)]^2-\Re B(s)^2+[\Im B(s)]^2
\nn \\
I_D(s)&=&2 s \Re A(s)\Im A(s) +2 \Re B(s) \Im B(s) \, ,
 \eea
and similarly for the function $\rho_v$.
\be \label{zdar2}
\rho_v(s)=\frac{-1}{\pi}\frac{\Im A(s) I_D(s)-\Im A(s)) R_D(s)}{R_D^2(s)+I_D^2(s)} \, .   
\ee

To get the solution of quark gap equation consistent with spectral representation one get the solution in two steps.
As a first step one solve the coupled system (\ref{fuc}) and (\ref{zdar},\ref{zdar2}) for  arbitrary initial  choice of four fixed numbers
$\Re B(\zeta), \Im B(\zeta), \Re A(\zeta), \Im A(\zeta)$ assuring the system is convergent, then one  changes gradually the imaginary parts 
in the  manner described bellow.

As the first,  two following functions 
\bea \label{cond1}
L_s&=&P.  \int_0^{\infty}d a\frac{\rho_s(a)}{p^2-a}
\nn \\
R_s&=&\frac{-\Im B(s) I_D(s)+\Re B(s) R_D(s)}{R_D^2(s)+I_D^2(s)}         
\eea  
should agree with each other. The equality obviously implies 
\be \label{equiv1}
L_s(s)=R_s(s)=\frac{1}{4} \Re Tr S(s)=\Re S_s(s)     
\ee
in this case. 

Similarly , defining for resulting solutions 
\bea \label{cond2}
L_v&=&P.  \int_0^{\infty}d a\frac{\rho_v(a)}{p^2-a}
\nn \\
R_v&=&\frac{-\Re A(s) R_D(s)+\Im A(s) I_D(s)}{R_D^2(s)+I_D^2(s)}         
\eea  
The equality obviously implies 
\be \label{equiv2}
L_v(s)=R_v(s)=\frac{1}{4p^2} \Re \not p Tr S(s)=\Re S_v(s) \, .     
\ee
These conditions are not used for purpose of  solving the system  (\ref{zdar})+(\ref{fuc})+(\ref{zdar2}).  Eqs. (\ref{equiv1}),(\ref{equiv2})  are obviously not fulfilled for arbitrary values  $ \Im B(\zeta)$ and $ \Im A(\zeta)$ and  are used as cross check instead.

All the system of equations  is then solved iteratively in two enclosed iteration cycles. The inner iteration cycles  solve the DSEs (\ref{zdar})+(\ref{fuc})+(\ref{zdar2})  iteratively. The external cycle scan imaginary parts of $A$ and $B$ at fixed points  and search   for their correct values by evaluating the norms
\bea
h(\sigma_v,\Im A)&=&\frac{\int (L_v(s)-R_v(s))^2 ds}{\int (L_v(s)+R_v(s))^2 ds}
\nn \\
h(\sigma_s,\Im B)&=&\frac{\int (L_s(s)-R_s(s))^2 ds}{\int (L_s(s)+R_s(s))^2 ds}
\eea
 
\begin{figure}
\centerline{\includegraphics[width=8.0cm]{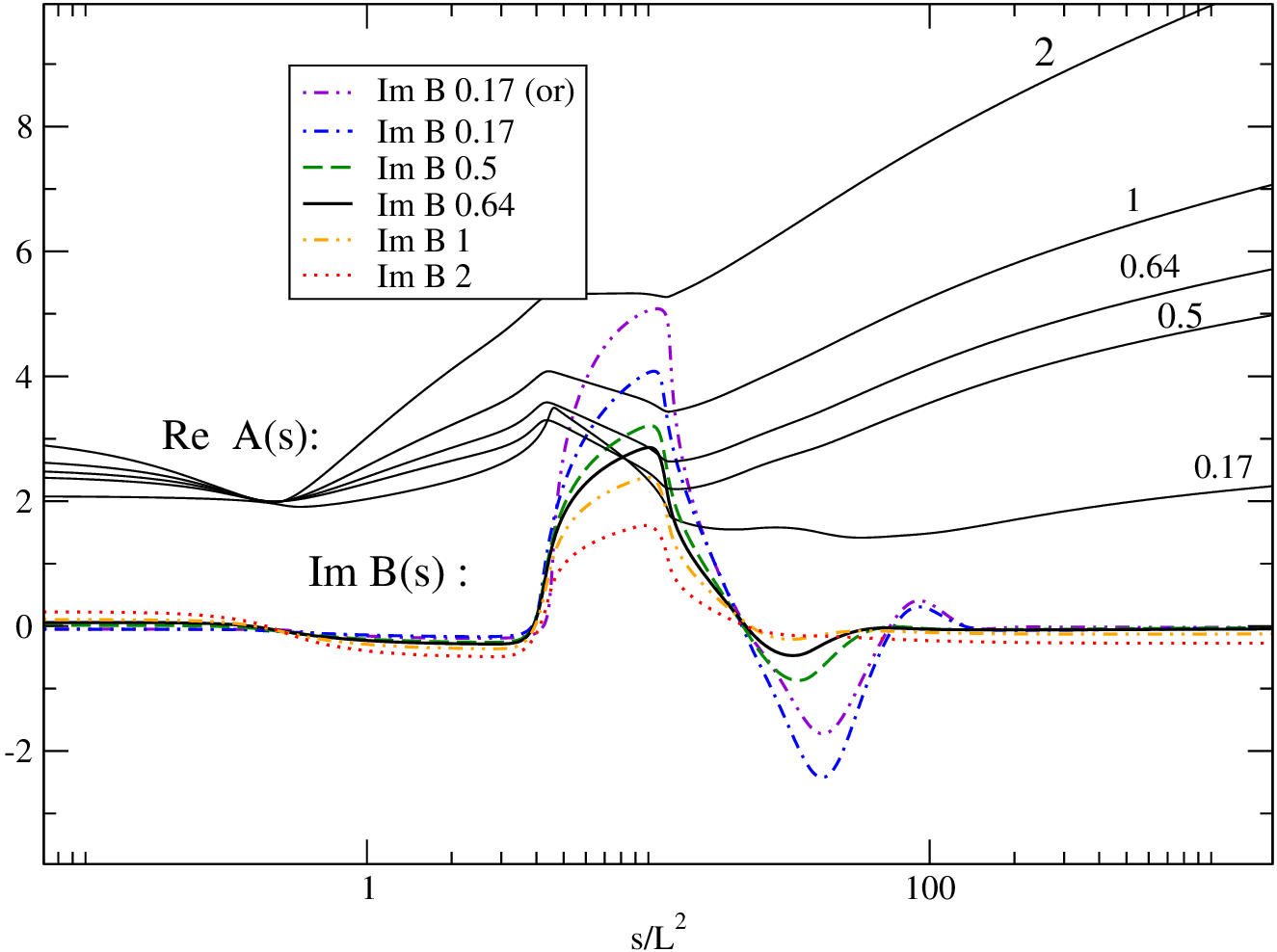}}
\caption{Solutions of the quark gap equation  as described in the  Appendix. Numbers refer various values of the product $g^2\xi$.}
\label{slava}
\end{figure}

The solution turns out to be  fast convergent and amazingly stable 
for almost any value of  gauge fixing parameters.
 Few seconds at recent computers are needed for ($\simeq 30$) iterations providing the stable solution for fixed masses and renormalization functions and therefore a correct ratio $\Im B(\zeta)/\Re B(\zeta)$   can be identified with an arbitrary precision. 
Examples of solutions obtained for different values of
the product  $g^2\xi$ is shown in the Fig. \ref{slava}. Up to the values $B(\zeta)$ and $A(\zeta)$ chosen such that $\Re B(\zeta)=1.0 L$ and 
$\Im A(\zeta)=2$ at $\zeta=0.1L^2$ all  other parameters values are identical to those referred in the main text and we show only the real part of renormalization function and the imaginary part of the function $B(s)$. Now, the  scale $L$ defines our units wherein $m_g^2=2L^2$ in this units.

The reader can recognize that the all above procedure remains a well know momentum renormalization scheme, here however performed at the timelike renormalization scale.
It can be regarded as, but  it should be stresses that  it has nothing to do with the renormalization of UV infinities presented in the theory,
noting that  the renormalization constants stay real since they were identified by imposing the renormalization conditions at spacelike t'Hooft renormalization scale  $\mu$ which is a must for theory with hermitian Lagrangian.

Providing the equalities (\ref{equiv1}) and (\ref{equiv2}) hold, the propagator evaluated at spacelike momentum  is equivalent  
to the solution of the  fermion DSE solved directly in the momentum Euclidean space. 
The uniqueness of analytical continuation ensures the statement exactly, while  mutual numerical comparison provides independent check 
(at least of numerical accuracy).

%%%%%%%%%%%%%%%%%%%%%%%%%%%%%%%%%%%%%%%%%%%%%%%%%%%%%%%%%%%%%%%%%%%%%%%

%
\end{document}